\documentclass[preprint,showpacs,preprintnumbers,showkeys,amsmath,amssymb]{revtex4}

\usepackage{graphicx}
\usepackage{dcolumn}
\usepackage{bm}

\begin{document}

\title{ Determination of the magnetic dipole moment of the rho meson using 4 pion electroproduction data}

\author{D. Garc\'{\i}a Gudi\~no and G. Toledo S\'anchez}
\affiliation{Instituto de F\'{\i}sica,  Universidad Nacional Aut\'onoma de M\'exico, AP 20-364,  M\'exico D.F. 01000, M\'exico}%

\date{\today}

\begin{abstract}
We determine the magnetic dipole moment of the rho meson using preliminary data from the BaBar Collaboration for the $e^{+} e^{-} \to \pi^+ \pi^- 2 \pi^0$ process, in the center of mass energy range from 0.9 to 2.2 GeV. We describe the $\gamma^* \to 4\pi$ vertex using a vector meson dominance model, including the intermediate resonance contributions relevant at these energies. We find that $\mu_\rho =  2.1 \pm 0.5 $ in $e/2 m_\rho$ units.
\end{abstract}

\maketitle
\section{Introduction}
To date there is no measurement of the magnetic dipole moment (MDM) of any vector meson \cite{pdg}. Their extremely short lifetimes ($\approx10^{-23}$ $s$) prevents experimentalists from applying standard MDM measurement techniques. For example, the spin precession technique \cite{spin} requires the determination of the spin polarisation of the particle before and after crossing a constant magnetic field region, a difficult task to accomplish for vector mesons. Alternatives to determine the MDM, by indirect means, invoke the fact that the radiation emitted from the vector meson carries out information of its electromagnetic structure \cite{radiative} and thus, provided the dominant electric radiation is known, the sub-leading MDM effect can be identified.
A determination of the spin 3/2 resonance $\Delta^{++}$, where experimental data on radiative $\pi^+ p$ scattering was available, has already made use of these ideas \cite{glcdelta}.\\
For the $\rho$ vector meson, studies of radiative decays of the form $\rho \rightarrow \pi\pi\gamma$, $\tau \rightarrow \nu \rho \gamma$ and $\tau \rightarrow \nu\pi\pi\gamma$ \cite{glcrho,glctau2pi}, which involve the radiative $\rho\rho\gamma$ vertex, have shown that there can be kinematical regions of the photon spectrum where electric charge contribution is suppressed, leaving the MDM effect as the leading one, thus offering a possibility to determine its value. Unfortunately, there are no experimental data for neither of these processes to confront with the theoretical description.\\ 
A variation of this approach is actually used to set the limits of the electromagnetic properties of the $W$ gauge boson, by looking at the triple gauge boson vertex ($WW\gamma$) in the $e^+ e^- \to jjl\nu$ process \cite{delphi} (where $j$, $l$ and $\nu$ denotes a jet a lepton and a neutrino, respectively). In similar way, we can notice that the $e^{+} e^{-} \to \pi^+ \pi^- 2 \pi^0$ process involves the $\rho\rho\gamma$ vertex in the channel where the pions are produced in pairs, through  $\rho$ meson resonant states, and therefore a determination of the electromagnetic properties of the $\rho$ meson may be possible.\\
In this work, we use preliminary data from the BaBar Collaboration \cite{babar4pi} for such process to determine the MDM of the $\rho$ meson. The $\gamma^* \to 4\pi$ vertex is modeled in the vector meson dominance (VMD) approach. We have include the channels involving the exchange of the $\pi$, $\omega$, $a_1$, $\sigma$, $f(980)$, $\rho$ and $\rho'$ mesons. The MDM is determined by fitting the experimental cross section data, while using other observables to fix all the remaining parameters.\\
This work is organised as follows: In section 2, we introduce the $\rho$ electromagnetic vertex and its multipole structure and the form of the propagator to account for the unstable feature. In section 3, we model the $e^{+} e^{-} \to \pi^+ \pi^- 2 \pi^0$ process exhibiting all the channels considered and discuss the corresponding approximations. In section 4, we compute the total cross section, exhibiting the different contributions and determine the MDM of the $\rho$ using the BaBar data. In section 5, as a byproduct, we compute the branching ratio for the $\rho \to \pi^+ \pi^- 2 \pi^0$ process and compare with the experimental value. In section 6 we discuss the results and present our conclusions.

\section{The electromagnetic vertex}
The electromagnetic vertex for  a vector particle (V) is defined by the electromagnetic current $<V(q_2,\eta)|J^\mu_{EM}(0)| V(q_1,\epsilon)>\equiv \eta_\nu^\dagger \epsilon_\lambda \Gamma^{\mu \nu \lambda}$, where $q_i$ are the momenta and $\epsilon$ and $\eta$ are the corresponding polarisation tensors.
The C, P and CP conserving electromagnetic vertex $\Gamma^{\mu \nu \lambda}$ can be decomposed into the following Lorentz structures 
\begin{eqnarray}
\Gamma^{\mu\nu \lambda} &=&
 \alpha(q^{2}) g^{\nu \lambda}(q_1 + q_2)^{\mu} + \beta(q^{2}) ( g^{\mu \nu} q^{\lambda} -  g^{\mu \lambda} q^{\nu})\nonumber\\
  &-& \frac{\gamma(q^{2})}{M_V^2} (q_1 + q_2)^{\mu} q^{\nu} q^{\lambda} 
  - q_1^\lambda g^{\mu \nu} -q_2^\nu g^{\mu \lambda},
\label{vertex}
\end{eqnarray}
where $\alpha(q^{2})$, $\beta(q^{2})$  and $\gamma(q^{2})$ are the electromagnetic form factors  \cite{hagiwara,nieves}. In the  static limit, the electromagnetic multipoles are identified as follows: $\mathcal{Q}_V = \alpha(0)$ is the electric charge ( in $e$ units), $\mu_V=\beta(0)$
 is the magnetic dipole moment (in $e/2 M_V$ units)  and the electric quadrupole is $X_{E_V} = 1-\beta(0)+2\gamma(0) $ (in $e/M_V^{2}$ units).
Another set of parameters to refer to the electromagnetic multipoles of spin-1 particles are $\kappa $ and $\lambda$  which are related to the previous ones by $\beta(0) \equiv1+\kappa+\lambda$ and $\gamma(0) \equiv \lambda$ \cite{hagiwara,nieves}. For instance, at tree level, the gauge structure of the standard model predicts   $\alpha(0) = 1$, $\beta(0) = 2$ and  $\gamma(0) =0 $ ($\kappa=1$ and $\lambda= 0$) for the $W$ gauge boson, corresponding to   $\mathcal{Q}_W=1$, $\mu_W=2$ and $X_{E_W} =-1$. These values are usually taken as a reference for vector mesons. However, since they are not linked to a gauge symmetry, they are rather expected to reflect the strong interaction dynamics among quarks. A plethora of effective approaches to QCD have been used to compute the MDM of  the light vector meson states  \cite{predictions}. The most representative is the $\rho$ meson, whose predictions for the MDM are found to lay in the region from 1.9 to 3 (in $e/2M_\rho$ units).\\
In addition, the proper theoretical description of the vector mesons requires the inclusion of its unstable feature (parameterized by the decay width, $\Gamma$) without breaking the electromagnetic gauge invariance.  The fermion loop scheme \cite{fermionloop}  and the boson loop scheme \cite{glctau2pi} (suitable for the $W$ and vector mesons respectively) succeed in this task by taking into account the absorptive contributions to the electromagnetic vertex and the propagator and the linearity of the Ward-Takahashi identity, which is fullfilled order by order in perturbation theory.
 In a previous work \cite{david:2010}, we have computed the correction to the multipoles of the $W$, $\rho$ and $K^*$ particles, exclusively from this fact, and found them to be relatively small. Moreover, these schemes are consistent with the complex-mass scheme \cite{complexmass} upon the renormalization of the vector field. Thus, it is well grounded to consider the above  expression for the electromagnetic vertex and for the vector meson propagator we use:
 \begin{equation}
 D^{\mu\nu}[q,V]=i\left(\frac{-g^{\mu\nu}+\frac{q^\mu q^\nu}{M_V-iM_V\Gamma} }{q^2-M_V^2+iM_V\Gamma}\right).
 \label{vectorpropagator}
 \end{equation}
This is the way we will consider the vector particles hereafter. The momentum dependence of the width will be used only for the $\rho$ meson.

\begin{equation}
\Gamma_\rho(q^2)=  \frac{\left(\sqrt{q^2}\right)^{-5}\left(\lambda \left[q^2,m_{\pi }^2,m_{\pi }^2\right]\right)^{3/2}}{m_{\rho }^{-5}\left(\lambda \left[m_{\rho }^2,m_{\pi }^2,m_{\pi }^2\right]\right)^{3/2}} \Gamma_\rho.
\end{equation}

where $\lambda [a, b, c] \equiv a^2+b^2+c^2-2 a b-2a c-2b c$.

\section{Modeling the $e^{+} e^{-} \to \pi^+ \pi^- 2 \pi^0$  process}
The $e^{+} e^{-} \to \pi^+ \pi^- 2 \pi^0$  process has been measured by several experiments in the low energy regime in a direct production from $e^+e^-$  \cite{ee4pi,snd4pi}, and preliminary  data is available from the BaBar collaboration \cite{babar4pi}, which uses the initial state radiation technique, in a wider energy range. A comparison of the total cross section data shows that SND and BaBar agree with each other for energies below 1.4 GeV \cite{4picompare}. Thus, for this study we made use of the BaBar data and consider the SND data for comparison purposes in the corresponding energy range.\\
The description of the data in the low energy range has been studied using effective models based on chiral symmetry and VMD \cite{eidelman,ecker,kuhn,lichard}. The approach we follow is based on VMD \cite{klz} which, by considering the relevant hadronic degrees of freedom in the energy range of our interest, is well suited to describe the process. Chiral symmetry models with resonances would be also applied \cite{xralresonance}.  In our case, the couplings are taken as effective constants determined from different observables, although in some cases relations can be drawn among them by invoking symmetry considerations.\\
Our notation for the process is: $ e^{+}(k_1) e^{-}(k_2) \to \pi^+(p_1) \pi^0(p_2) \pi^-(p_3) \pi^0(p_4)$, in parenthesis are the corresponding 4-momenta. The total amplitude can be written as:
\begin{equation}
\mathcal{M}=\frac{-ie }{(k_1+k_2)^2} l^\mu h_\mu(p_1, p_2 ,p_3, p_4) ,
\end{equation}
where the leptonic current $l^{\mu} \equiv  \bar{v}(k_2) \gamma^\mu u(k_1)$ is common to all the channels, and  $h_\mu$ represents the four pion electromagnetic current. This last must fulfill the Bose-Einstein symmetry, by the interchange of the neutral pions
\begin{equation}
h_\mu (p_1, p_2 ,p_3, p_4)=h_\mu(p_1, p_4 ,p_3, p_2),
\end{equation}
and $C$ invariance, by the interchange of the charged pions
\begin{equation}
h_\mu (p_1, p_2 ,p_3, p_4)=- h_\mu (p_3, p_2 ,p_1, p_4).
\end{equation}
Thus, the total contribution  can be written as the sum of the four possible momenta configurations, represented by a reduced amplitude $\mathcal{M}_{r \mu}$ no longer constrained by such symmetries \cite{ecker}:
\begin{eqnarray}
h_\mu (p_1, p_2 ,p_3, p_4) &=& \mathcal{M}_{r \mu} (p_1, p_2 ,p_3, p_4) + \mathcal{M}_{r \mu} (p_1, p_4 ,p_3, p_2)  \nonumber \\
&& - \mathcal{M}_{r \mu} (p_3, p_2 ,p_1, p_4) - \mathcal{M}_{r \mu} (p_3, p_4 ,p_1, p_2).
\label{ampcanaltot}
\end{eqnarray}

\begin{figure}
\begin{center}
\includegraphics[scale=0.5]{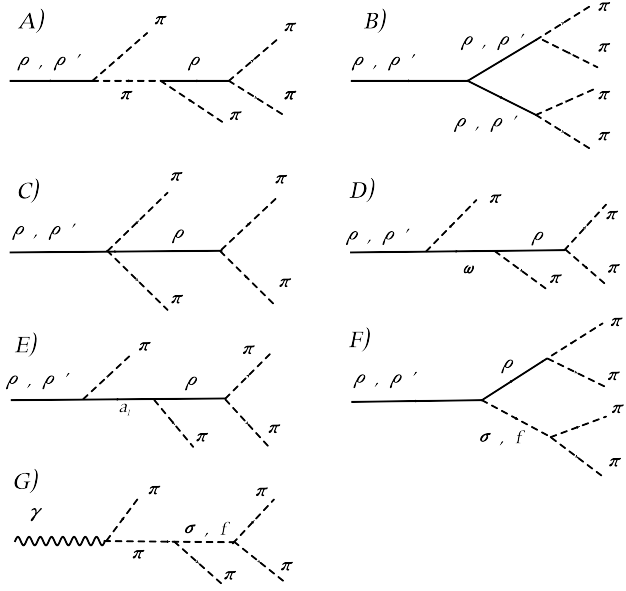}
\end{center}
\caption{Generic channels relevant for the description of the $e^{+} e^{-} \to \pi^+ \pi^- 2 \pi^0$  process. The total diagrams are obtained by applying Bose symmetry and Charge conjugation.}
\label{procesos}
\end{figure}

To model the four pion electromagnetic current, we consider the channels including the exchange of the $\pi$, $\omega$, $a_1$, $\sigma$, $f(980)$, $\rho$ and $\rho'(1450)$  mesons, as shown in Figure \ref{procesos}. The energy range to be described goes from threshold  up to 2.2 GeV. Thus, we have seven generic channels, each one accounting for several specific diagrams, corresponding to the allowed permutations of the momenta due to Bose-Einstein symmetry and charge conjugation. We now proceed to discuss each one in detail.

\subsection{Channel A}
The diagram (A) shown in Figure \ref{procesos}, corresponds to the case where we have a $\pi$ and $\rho$ intermediate states, and the initial resonant state coupled to the leptonic current are both a $\rho$ and $\rho'$ mesons.
The interaction Lagrangian involving vectors (V), pseudoscalars (P) and the photon (A) is: 
\begin{eqnarray}
{\cal L}&=& g_{VPP} \epsilon_{abc} V_\mu^a P^b \partial^\mu P^c 
+\frac{e m_V^2}{g_V}V_\mu A^\mu 
\label{lagrangian}
\end{eqnarray}
where $g_{VPP}$ and $g_V$ are effective coupling constants.
 The reduced amplitude for such contribution considering the $\rho$ intermediate state is then given by:
\begin{eqnarray}
\mathcal{M}_{A \mu} & = & e\left( \frac{ m_{\rho}^2  g_{\rho \pi \pi }^3}{g_\rho }\right) 
 D_{\mu\lambda}\left[q,\rho \right] \nonumber\\
&&  (q-2p_1)^\lambda
D\left[q-p_1,\pi \right] (q-r_{12})^\nu D_{\nu\delta}\left[s_{34},\rho \right] r_{34}^\delta,
\label{mproared}
\end{eqnarray}
where $q \equiv k_1+k_2=p_1+p_2+p_3+p_4$, 
$s_{ij} \equiv  p_i + p_j$, 
$r_{ij} \equiv  p_i - p_j$. $D_{\mu \nu}[q, V]$ is the vector meson propagator as given in Eqn. \ref{vectorpropagator} and $D[q, P]=i/(q^2-m_P^2)$ is the pseudoscalar propagator, the arguments denoting the four momentum and the corresponding particle.
A similar expression is obtained for the case when the photon couples to the $\rho'$, which  is considered to have a relative phase respect to the $\rho$ of $180^0$, this has been shown to be the case in \cite{ecker,kuhn} and will be used without further mention in the following diagrams. We have verified that this value is favored in the analysis of the cross section (see the corresponding section below).
Given the scarce information on the $\rho'$ decay modes, we  asume the following combination of couplings constants for the $\rho'$ to be the same as for the $\rho$ . 
\begin{equation}
 \dfrac{m^2_{\rho '}}{g_{\rho '}} g_{\rho ' \pi \pi} = \dfrac{m^2_{\rho}}{g_\rho} g_{\rho \pi \pi}.
 \label{consacopaprox}
\end{equation}
The idea behind this assumptions is to resemble typical VMD relations, expecting the particularities of the radial excitation properties drops out when considering the ratios. Implications of the deviation from this assumption are explored in the analysis.  
Thus, the parameters involved in this channel are  $g_{\rho\pi\pi}=5.96 \pm 0.02$  and $g_{\rho}=4.96 \pm0.02$ which are determined from the $\rho \to \pi\pi $ and $\rho \to e^+ e^-$ decay respectively.

\subsection{Channel B}
This channel corresponds to diagram (B) in Figure \ref{procesos}. The extraction of the MDM of the $\rho$ meson is based on the existence of the $\rho-\rho-\gamma$ vertex. An energy range up to 2.2 GeV allows the photon to couple to intermediate states like the $\rho$ and $\rho'$ mesons, which then couple to the $\rho$ pair. Thus, the structure of such vertices is similar to the electromagnetic vertex but  with different global constants accounting for the strong process. It is possible to identify such constants with the $ g_{\rho \pi \pi }$ and $ g_{\rho' \pi \pi }$ couplings respectively, considering the $\rho$ as an SU(2) gauge boson and generalizing to the $\rho'$. These effects are not arbitrary as they are linked by the electric charge form factor (see subsection H, below). 

Let us illustrate the form of the reduced amplitude, considering only the $\rho$ meson triple vertex:
\begin{equation}
\mathcal{M}_{B \mu} = e\left( \frac{ m_{\rho}^2  g_{\rho \pi \pi }^3}{g_\rho }\right)
 D_{\mu \nu}\left[q, \rho\right] \Gamma^{\nu\theta\tau}
  D_{\theta \delta}\left[s_{12} , \rho\right] D_{\tau \eta}\left[s_{34} , \rho\right] r_{21}^\delta r_{34}^\eta,
\label{mprob}
\end{equation}	
the only free parameters are those involved in the electromagnetic vertex $\Gamma^{\nu\theta\tau}$ Eqn. (\ref{vertex}). Namely, the $\beta$ and $\gamma$ parameters, since the electric charge is fixed.\\
We have similar expressions for the case when the $\rho'$ is involved, which are added under the same considerations mentioned in the previous section. For the $\rho'$ triple vector meson vertex, we take the same structure and coupling as for the $\rho$ case, this assumption has been found to be appealing \cite{kuhn}, and its implications in our analysis will be discussed below.

 \subsection{Channel C}
The diagram (C) in Figure \ref{procesos} includes a $\rho\rho\pi\pi$ contact term, with the subsequent decay of the $\rho$ into two pions. The contact coupling is fixed by requiring gauge invariance of the sum of the (A), (B) and (C) amplitudes:
 \begin{equation}
 q^\mu(h_{ A \mu}+h_{B \mu}+h_{C \mu})=0. 
\end{equation}
 The reduced amplitude can be written as:
\begin{equation}
\mathcal{M}_{C \mu}  =  \left(\frac{ e m_{\rho }^2}{g_{\rho }} g_{\rho \pi \pi } \right)
D_{\mu \nu}[q, \rho] g_{\rho \rho \pi \pi } T^{\nu \delta }
D_{\delta \gamma}[q, \rho] r_{34}^\gamma 
\label{mprocprop}
\end{equation}
where the $g_{\rho \rho \pi \pi }$ coupling and the  $T^{\nu \delta }$ tensor are fixed by the gauge invariance condition.
Note that the gauge invariance condition is applied  for every particular form of the corresponding contribution from Bose-Einstein symmetry and charge conjugation, each one producing an equivalent form for the contact contribution.
  
  \subsection{Channel D}
\begin{figure}
\begin{center}
\includegraphics[scale=0.35]{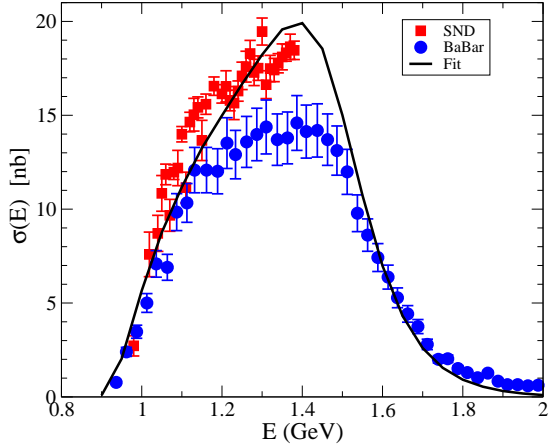}
\end{center}
\caption{Cross section for the $e^{+} e^{-} \to \pi^+ \pi^- 2 \pi^0$ process due exclusively to the $\omega$ channel (D). Data from BaBar \cite{babar4pi} and SND \cite{snd4pi} are fitted to obtain the $g_{\omega \rho' \pi}$ coupling.}
\label{gwrpp}
\end{figure}

This channel corresponds to the contribution of the $\omega$ and $\rho$ meson intermediate states as depicted in the diagram (D) of Figure \ref{procesos}. The $\omega-\rho (\rho')-\pi$ coupling is given by the following interaction Lagrangian:
\begin{eqnarray}
{\cal L}_{\omega}&=&  g_{\omega\rho\pi}\delta_{ab}\epsilon^{\mu\nu\lambda\sigma}\partial_\mu \omega_\nu \partial_\lambda \rho_\sigma^a  \pi^b . 
\label{omega}
\end{eqnarray}
The corresponding reduced amplitude is given by:
\begin{eqnarray}
\mathcal{M}_{D \mu} & = & -e\frac{m_{\rho }^2 g_{\rho \pi \pi } g_{\omega \rho \pi } }{g_{\rho }} g_{\omega \rho \pi } D_{\mu\nu}[q,\rho ]\nonumber\\
&&
\epsilon^{\xi \nu \gamma\alpha}q_{\xi }p_{2\gamma}
D_{\alpha\eta}\left[q-p_2,\omega \right]
\epsilon^{\phi\eta\sigma\theta} (q-p_2)_\phi  s_{{34}\sigma}
 D_{\theta\lambda}\left[s34,\rho \right] r_{34}^\lambda
\label{ampd}
\end{eqnarray}	
where the $\rho$ intermediate state is threefold, as it can be charged and neutral. An analogous expression is obtained for the $\rho'$. This and the remaining diagrams are gauge invariant by themselves.
The  required couplings are $g_{\omega \rho \pi}$ and $g_{\omega \rho' \pi}$. The first one has been determined to be  $g_{\omega \rho \pi}=14.7 \pm 0.1$ GeV$^{-1}$ from an analysis of vector mesons radiative decays, the $\omega \rightarrow 3\pi$ decay width and the $e^+e^- \to 3\pi$ cross section \cite{IJMPA}. To find the $g_{\omega \rho' \pi}$ coupling we fit the SND \cite{snd4pi} and BaBar \cite{babar4pi} data for this channel  (E$<$2 GeV), with this coupling as the only free parameter.  Requiring the sum of the $\chi^2$ of the fit to each data set to lead to a minimum ($\chi^2_{SND}/DoF$ +$\chi^2_{BABAR}/DoF$), we get $g_{\omega \rho' \pi}=10.8 \pm 0.6$ GeV$^{-1}$, where the error bar accounts for the difference to fit both data sets individually. 
 In Figure \ref{gwrpp}, we show the best fit  under this criterium, with the assumption of a relative phase of $180^0$ between the $\rho$ and $\rho'$ contributions. This assumption was explored and its role found to be relatively significant in the energy region above $M_{\rho'}$; as we will show later, in that region the contribution from this channel to the total process becomes subdominant and therefore the effect of the phase itself is mild.
 
\subsection{Channel E}
This channel involves the $a_1$ axial vector meson and the $\rho$ ($\rho'$) meson intermediate states as depicted in the diagram (E) of Figure \ref{procesos}. The simplest form of the effective Lagrangian for the $a_1(q) - \rho(k)-\pi(p) $ strong interaction (where $q$, $k$ and $p$ are the corresponding four momenta) is taken to be  \cite{a1isgur}:
\begin{eqnarray}
{\cal L}_{a_1}&=&  2 g_{a_1\rho\pi} (\rho_{\mu} a_1^\mu- \frac{\partial_\nu \rho^\mu \partial_\mu a_1^\nu}{k\cdot q})
\label{a1}
\end{eqnarray}
The reduced amplitude is then:
\begin{eqnarray}
\mathcal{M}_{E \mu} & = & \left(\frac{e m_{\rho }^2}{g_{\rho }} 4 g_{a\rho \pi }^2 g_{\rho \pi \pi }  \right) 
 D_{\mu\nu}[q,\rho ]
(q\cdot (q-p_1) g^{\nu\lambda}-q^\nu (q-p_1)^\lambda)\nonumber\\
&&D_{\lambda \alpha}[q-p_1,a_1 ]
(s_{34}\cdot (q-p_1) g^{\alpha\beta}-s_{34}^\alpha (q-p_1)^\beta)
 D_{\beta\gamma}[s_{34},\rho ] r_{34}^\gamma,
\label{ampproe}
\end{eqnarray}	
here the width of the $a_1$ is taken as a constant and the coupling $g_{a_1 \rho \pi}=3.25 \pm 0.3$ GeV$^{-1}$  is determined from the $a_1 \to \rho \pi$ decay. The corresponding coupling to $\rho'$ is taken to be the same. As we will show later, this channel is very suppressed in the whole region of study and deviations from this assumption are expected to have a very small effect. 

  \subsection{Channel F}

This channel  involves the $\rho$ and a scalar particle intermediate states as depicted in the diagram (F) of Figure \ref{procesos}. We consider the scalar to be both $\sigma(600)$ and $f_0(980)$. The interaction between the vector (V) and pseudoscalar (P) particles with the scalar (S) are parameterised by:
\begin{eqnarray}
{\cal L}_{S}&=&  g_{V_1V_2 S} V_{1\mu} V_2^\nu S+ g_{SP_1P_2}SP_1P_2
\label{scalar}
\end{eqnarray}
where  $g_{V_1V_2S}$ and $g_{SP_1P_2}$ are the effective coupling constants.
The reduced amplitude takes the following form:
\begin{eqnarray}
\mathcal{M}_{r F \mu} & = & \left(-\frac{i e m_{\rho }^2}{g_{\rho }} \right)g_{\rho \rho \sigma } g_{\rho \pi \pi } g_{\sigma \pi \pi }
D_{\mu\nu}[q,\rho ]
 D[s_{24},\sigma] D_{\nu\lambda}[q-s_{24},\rho ]r_{13}^{\alpha}.
\label{ampprof}
\end{eqnarray}

 The VMD relation $g_{\rho \rho \sigma}= -(e/g_\rho)g_{\rho \sigma \gamma}$ allows to determine $g_{\rho \rho \sigma}$, where $g_{\rho \sigma \gamma}= 0.63\pm 0.15$ GeV$^{-1}$ is determined from the $\rho \to \sigma \gamma$ decay. The coupling $g_{\sigma\pi\pi}=3.7 \pm 1.6$ GeV is determined from the $\sigma \to \pi\pi $ decay. For the f(980) we use the same coupling constants. The effect of the large uncertainties will be reflected in the low energy regime of the cross section. 

  \subsection{Channel G}

We consider a non-resonant $\rho$ and $\rho'$ channel, as represented in the diagram (G), including an intermediate scalar particle, that can be both the $\sigma$ and $f(980)$.  As we pointed out above, the information on the couplings and parameters of the scalars are not well determined, producing a strong source of uncertainties in the low energy regime. As we will show below, this lack of precision do not affect the region of our interest to determine the MDM of the  $\rho$ meson.\\
The reduced amplitude is given by:
\begin{eqnarray}
\mathcal{M}_{r G \mu} = 2 e g_{\sigma \pi \pi }^2
 D[q-p_1,\pi ] D[s_{24},\sigma ]
\left(p_{1\mu}- \frac{q\cdot p_1}{q\cdot p_3}p_{3\mu}\right)
\label{ampprog}
\end{eqnarray}
which has been built to be gauge invariant by itself. Note that, although we have used the same notation for the pseudoscalar and  scalar propagators, this last includes the corresponding decay width by replacing $m^2 \to m^2-im\Gamma$.

\subsection{Electric charge form factor}

The electromagnetic structure of the $\rho$ meson as a function of the momentum is accomplished by the inclusion of the $\rho$ and $\rho'$ resonances coupled to the photon, such that the electric charge form factor is written as
\begin{equation}
F_{\rho }\left(q^2\right) = \frac{g_{\rho  \pi \pi }m_{\rho }^2}{g_{\rho }}\sum_{j=\rho,\rho'}
\frac{1}{q^2-m_j^2+ i m_j \Gamma _j}.
\label{fformarec}
\end{equation}
Here we have made use of two previous assumptions, Eqn. (\ref{consacopaprox}) 
and the sum is made using a relative phase of $180^0$. The condition $F_{\rho }(q^2\rightarrow 0) \to -1$ imposes the relation between the remaining couplings to be:
\begin{equation}
\frac{g_{\rho \pi \pi }}{g_{\rho }}\left(\frac{m_{\rho '}^2-m_{\rho }^2}{m_{\rho '}^2}\right)=1.
\label{relff}
\end{equation}
Using the corresponding numerical values we find it to be equal to 0.86. Thus, we normalize the electromagnetic form factor to this value to have it properly defined.
Note that in our calculation the $\rho$ width is momentum dependent while the $\rho'$ width is taken as a constant and the limit taken above may look inconsistent. By comparing the Breit-Wigner distribution for the $\rho'$ when including/excluding the momentum dependence down to the 4-pion threshold we have verified that they are consistent with each other. Thus, it is justified the use of a constant width for the calculation and a momentum dependent width when taking the zero momentum limit.

\begin{figure}
\begin{center}
\includegraphics[scale=0.35]{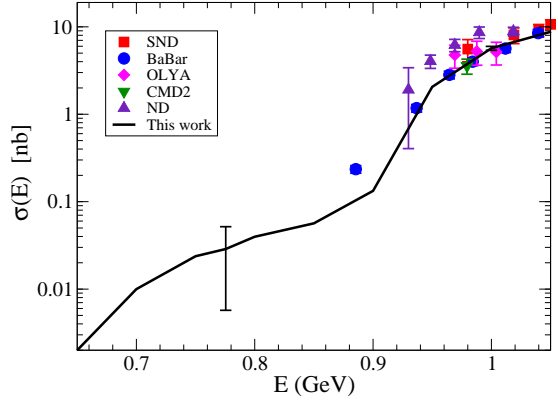}
\end{center}
\caption{Total cross section  $e^{+} e^{-} \to \pi^+ \pi^- 2 \pi^0$ in the energy region from threshold to 1.4 GeV, compared to several experimental data: SND \cite{snd4pi}, BaBar \cite{babar4pi}, OLYA, CMD2 and ND \cite{ee4pi} }
\label{lowcross}
\end{figure}	

\section{ Cross section of $e^{+} e^{-} \to \pi^+ \pi^- 2 \pi^0$.}
Now, we proceed to compute the cross section
\begin{equation}
d \sigma = \frac{(2\pi )^4}{4k_1.k_2} \vert \mathcal{\bar M} \vert^2 d\Phi,
\label{secefic}
\end{equation}
where $d\Phi$ is the 4-body phase space and $\vert \mathcal{\bar M} \vert^2$ is the averaged squared amplitude given by  
\begin{eqnarray}
\vert \mathcal{\bar M} \vert^2 = \dfrac{e^2}{4 q^4}  l^{\mu \nu} h_{\mu \nu};
\label{amptot}
\end{eqnarray}
The leptonic contribution after sum over polarizations is: 
$l^{\mu \nu} = k^{\mu}_1 k^{\nu}_2 + k^{\nu}_1 k^{\mu}_2 - k_1 \cdot k_2\: \: g^{\mu \nu}$,
and $h_{\mu \nu}$ is the square of the four pion current for all the above channels. The integration is performed numerically using a Fortran code and the Vegas subroutine. The kinematical phase space configuration is implemented as described in Ref. \cite{kumar}.

In Figure \ref{lowcross}, we plot the cross section in the low energy region (below 1.1 GeV). 
The result from our model (solid line) is compared to experimental results from SND \cite{snd4pi}, BaBar \cite{babar4pi}, OLYA, CMD2 and ND \cite{ee4pi} (symbols), which are properly described. Our study shows that, in this region, the cross section is dominated by the $\omega$ and $\sigma$ channels (D) and (G), consistent with what has been found in previous analysis \cite{kuhn}. Error bars are displayed at some representative points and are dominated by the $\sigma(600)$ parameters. In this region there is no effect due to variations of the parameters of channel (B).

\begin{figure}
\begin{center}
\includegraphics[scale=0.32]{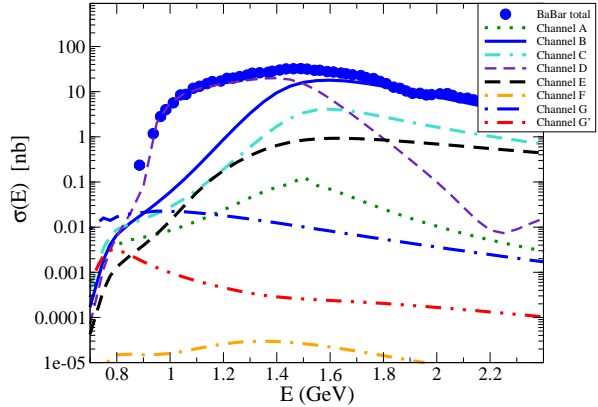}
\end{center}
\caption{Individual channel contributions to the total cross section for $e^{+} e^{-} \to \pi^+ \pi^- 2 \pi^0$ and the BaBar\cite{babar4pi} experimental data.}
\label{allchannels}
\end{figure}	

In Figure \ref{allchannels}, we have plotted all the individual channels contributing to the total cross section and the preliminary experimental data from BaBar. Each channel includes the full reduced amplitudes for $\rho$ and $\rho'$ and their corresponding interferences, which are the dominant ones.  The interferences among different channels are relatively smaller and not shown but accounted in the analysis. 
Error bars are not displayed for the sake of clarity, but we will comment on them to give a glimpse of their importance. We observe that the energy region below 1.4 GeV is dominated by the $\omega$ channel (D), the subsequent decrease of this channel is associated to the interference between the $\rho$ and $\rho'$. The uncertainties for this channel come mainly from the combination of the error bars of the coupling constants and are in the range between 8\% and 15\%.
 Above 1.4 GeV the (A), (B) and (C) channels, which are linked by gauge invariance, increase their effects and eventually the channel (B) (here displayed for $\beta=2$ and $\gamma=0$) surpass the $\omega$ contribution, becoming the leading one, followed by channel (C). 
 This observation justify our earlier statement about the mild effect of the relative phase between the $\rho$ and $\rho'$ used in channel (D) in the region where the channel (B) becomes important.
 Channels (A) and (C) have uncertainties of 15\%,  dominated by the uncertainties of the mass and widths of the particles involved.
 The $a_1$ channel (E) also increases its effect but still below channel (C). 
This justify our assumption that the relation between the  couplings of $a_1$ to  $\rho$ and $\rho'$ used in channel (E), although not well grounded experimentally, should play no important role in this analysis.  We have associated an overall 16\% uncertainty to this channel, based on the mass, width and estimates of the effect from deviations of the couplings relation of $\pm$10\%.
 Channel (F) is far below all contributions in the whole region under consideration. The effect from the $f(980)$ is shown on the same basis as the $\sigma$  channel (G) (denoted by G') and is also relatively small. 
 The uncertainties associated to these channels can be as large as 70\% (F), 90\%(G') and 200\% (G) but, given their small relative size, they have no impact in the outcome of the analysis.
\begin{figure}
\begin{center}
\includegraphics[scale=0.35]{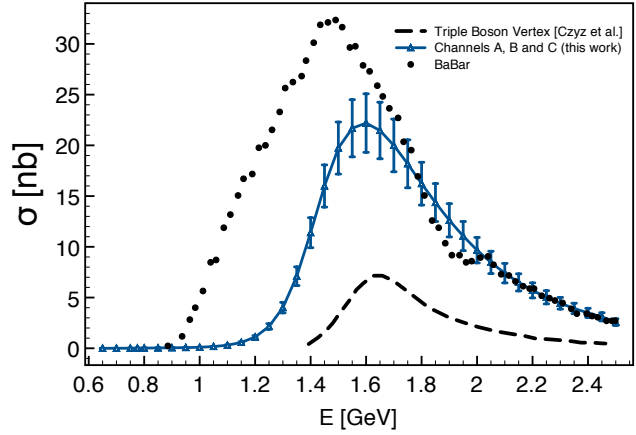}
\end{center}
\caption{Contribution to the total cross section (BaBar \cite{babar4pi}, dotted line) from the sum of channels (A), (B) and (C)  considered in this work (solid line) and the triple vector boson vertex contribution obtained in \cite{kuhn} (long-dashed line). }
\label{nosvskuhn}
\end{figure}		
Note that, in a broader sense, a so-called $\rho$ channel, can not be restricted to channel (B) alone. As we have described above, it is linked by gauge invariance to channels (A) and (C), and a proper identification must be made through the quantification of the sum of these channels. In Figure \ref{nosvskuhn}, we show the sum of these contributions (solid line), which is very close to a non-omega contribution to the cross section. We also plot the result obtained in \cite{kuhn} for the triple vector boson vertex contribution (dashed line), whose structure is similar to our (B) plus (C) channels but with different coupling constants, obtained from a global fit.

\begin{figure}
\begin{center}
\includegraphics[scale=0.3]{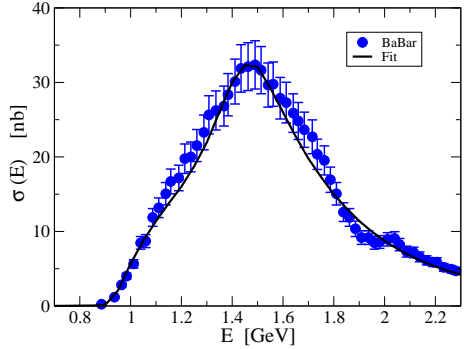}
\end{center}
\caption{Fit to the BaBar data \cite{babar4pi} (solid line) for the total cross section $e^{+} e^{-} \to \pi^+ \pi^- 2 \pi^0$.}
\label{betacross}
\end{figure}

In Figure \ref{betacross} we show the total cross section data from the preliminary analysis of BaBar\cite{babar4pi}, where we have assigned a 10\% systematic error bar (symbols).
Provided all the parameters involved in our description are determined from other observables, except the ones of channel (B), namely $\beta$ and $\gamma$ in the electromagnetic vertex, we performed a fit considering $\gamma =0$ and leaving $\beta$ as the only free parameter. The solid line represents the best fit, corresponding to $\mu_\rho=2.1$. For energies above 1.6 GeV there are structures that are not well captured by the current description, but they are relatively small compared to the effects that the MDM can produce, as shown in the figure. 
The fit considering $\beta$ and $\gamma$ as  free parameters favors the same $\beta$ and is loosely affected by $\gamma$ at the end region under consideration. Since in this region we expect to have effects from other resonances like the $\omega(1420)$ and $\rho(1700)$, not included in the current description, we restrict ourselves to the case $\gamma=0$.

To determine the total cross section error bars, we take into account the combined uncertainties coming from the couplings of the different channels, assumed as no correlated.
 In addition, for channel (B) we explored the role of the model assumption regarding the global combination of couplings and mass of the $\rho'$ taken to be similar to the $\rho$ (see Eqn. \ref{consacopaprox}). It was found to be consistent with data for upto a $\pm$10\% deviation respect to the $\rho$ case.
We finally determined the $\beta$ parameter error bar considering it as the responsible of the total uncertainties of the cross section. In addition, to account for the model dependence, we have added a 20\% error bar (added in quadrature). Thus, we obtain that the MDM of the $\rho$ meson is  
\begin{equation}
\mu_\rho = 2.1 \pm 0.5 \ \ \text{in} \ (e/2 m_\rho) \ \text{units}.
\label{resfinbeta}
\end{equation}

\section{$\rho^0 \to \pi^+ \pi^- 2 \pi^0$ branching ratio}

The branching ratio of the $\rho \to \pi^+ \pi^- 2 \pi^0$ decay can be computed by using the value of the cross section at the pole of the $\rho$ meson, without considering the $\rho'$ and the non-resonant channel, as follows \cite{ecker}:
\begin{equation}
 BR(\rho^0 \to \pi^+ \pi^- 2 \pi^0) = \frac{ m_{\rho }^2 \sigma \left(e^+e^-\to  \pi^+ \pi^- 2 \pi^0 \right)|_{E= m_{\rho }}}{12 \pi  BR \left(\rho ^0\to  e^+e^-\right)},
 \end{equation}

where $BR \left(\rho ^0 \to  e^+e^-\right)=4.72 \pm 0.05 \times 10^{-5}$  \cite{pdg}. 
The contribution to the total cross section at the $\rho$ pole is mainly affected by channels (A), (B), (C) and (D) and their interferences. The rest of the contributions have a relatively small contribution.
The cross section we obtain corresponds to a branching ratio of
\begin{equation}
BR(\rho^0 \to \pi^+ \pi^- 2 \pi^0 ) = 1.7 \pm 0.6 \times 10^{-5} ,
\label{br2}
\end{equation}
which is in agreement with the experimental value $BR(\rho^0 \to \pi^+ \pi^- 2 \pi^0 ) = 1.6 \pm 0.8 \times 10^{-5}$ \cite{snd4pi,pdg}. The error bar is taken with the same considerations as for the MDM.   To illustrate the mild dependence of the branching ratio on the value of the MDM, in Table \ref{table1}, we show the central value of the branching ratio for a wide range of values of the MDM, which still in agreement with the experimental result within the current uncertainties.

\begin{table}
\begin{center}
\begin{tabular}{|c|c|}
\hline
MDM &$BR(\rho^0 \to \pi^+ \pi^- 2 \pi^0 )$\\
\hline
1.2 &1.5 $\times 10^{-5}$\\
2.1 &1.7  $\times 10^{-5}$\\
3.5 & 2.1  $ \times10^{-5}$\\
\hline
\end{tabular}
\end{center}
\caption{$\rho^0 \to \pi^+ \pi^- 2 \pi^0$ branching ratio for a set of values of the $\rho$ MDM.}
\label{table1}
\end{table}
\section{Conclusions}
We have determined the $\rho$ meson MDM from the $e^+ e^- \to \pi^+ \pi^- 2 \pi^0$ cross section, using preliminary data from the Babar Collaboration. We modeled the $\gamma^* \to 4 \pi$  vertex considering the exchange of the $\pi$, $\omega$, $a_1$, $\sigma$, $f(980)$, $\rho$ and $\rho'$ mesons. The behavior of the cross section below 1.4 GeV is dominated by the properties of the intermediate scalar and the $\omega$  meson. The channel that contains the electromagnetic vector meson vertex becomes relevant for energies between 1.5 and 2.2 GeV, while the remaining channels are always subdominant.
We have found that the best fit implies a value for the MDM of the $\rho$ meson of $\mu_\rho=  2.1 \pm 0.5 $ in ($e/2 m_\rho)$ units.
The quoted error bar takes into account the uncertainties coming from the couplings of the different channels and model assumptions. Definite data on this process and detailed information on the $\rho'$ meson properties will be very useful for a more refined analysis to support a definite value. Preliminary data from SND \cite{SNDpreliminary} seems to be in agreement with the one from BaBar.
Additionally, we computed the branching ratio  for the $\rho \to \pi^+ \pi^- 2 \pi^0$ decay, which was found to be consistent with the experimental value, exhibiting a mild dependence on the $\rho$ meson MDM.\\
 We would like to conclude by stating that a long standing problem on the determination of the MDM of the $\rho$ meson can be addressed, and with our analysis we have provided a first insight on it.

\begin{acknowledgments}
We acknowledge the support of CONACyT-M\'exico under grant 128534, and DGAPA-PAPIIT UNAM, under grants IB101012 and IN106913. We thank Dr. G. L\'opez Castro and Dr. Jens Erler for very useful observations.
\end{acknowledgments}

\end{document}